# Architectural Challenges of Nomadic Networks in 6G

Daniel Lindenschmitt[1,*], Benedikt Veith[2,†], Khurshid Alam[3,†], Ainur Daurembekova[4,*], Michael Gundall[5,†], Mohammad Asif Habibi[6,*], Bin Han[7,*], Dennis Krummacker[8,†], Philipp Rosemann[9,*], and Hans D. Schotten[10,*†]

*University of Kaiserslautern (RPTU), Germany.

†German Research Center for Artificial Intelligence (DFKI), Germany.

[1]daniel.lindenschmitt@rptu.de; [2]benedikt.veith@dkfi.de; [3]khurshid.alam@dfki.de; [4]ainur.daurembekova@rptu.de; [5]michael.gundall@dfki.de; [6]m.asif@rptu.de; [7]bin.han@rptu.de; [8]dennis.krummacker@dfki.de; [9]philipp.rosemann@rptu.de; [10]schotten@rptu.de

## Abstract

This paper examines architectural challenges and opportunities arising from Nomadic Networks in the context of emerging 6G research. Nomadic networks are proposed as a solution to the limitations of stationary communication networks, providing enhanced connectivity for dynamic and mobile environments, such as large outdoor events, emergency situations and mobile industrial applications. The key requirements for nomadic networks are outlined, including functional split within the Radio Access Network and robust backhauling solutions. It also addresses the complexity of managing network components, ensuring interoperability with existing systems and maintaining stakeholder trust. A comprehensive architectural framework for Nomadic Networks in 6G is proposed. Different deployment scenarios for Nomadic Networks are investigated, including spawned, steered, and wandering Radio Access Networks as well as integrated, migrated and donated Core Networks. By introducing Nomadic-Network-as-a-Service and a related orchestration framework, the potential for flexible and scalable network management is emphasized. By addressing the architectural challenges, the paper provides a path for the successful implementation of Nomadic Networks towards more adaptable and flexible 6G networks that can meet the evolving needs of multiple sectors.

*Preprint.*

## 1 Introduction

Dynamic and self-organizing networks will play a critical role in upcoming technologies such as the 6G mobile communications standard. This change is driven by increasing demands from various sectors, including industry, manufacturing, agriculture, and the public sector, each requiring more specific functionalities. The establishment of NonPublic Networks (NPNs) within the current 5G standard has laid a crucial foundation, allowing independent operations within certain frequencies and localized areas, particularly for Internet of Things (IoT) applications. This concept is similar to the Citizens Broadband Radio Service (CBRS) in the 3550-3700 MHz frequency range, where the Spectrum Access System manages frequency use to protect incumbents and priority access licensees.

In 5G, mobile networks are constrained by stationary operation. Introducing a new type of mobile networks, so called Nomadic Networks (NNs), can unlock new application fields and allow a more flexible and comprehensive use of 6G technology. Imagine a bustling outdoor event: a concert, a sports match, or emergency scenarios. Here, the base station is not a monolithic entity; it is a dynamic ensemble. The dynamicity of networks, particularly in dynamic environments, plays a crucial role in establishing flexible and adaptive network configurations. This adaptability enables efficient allocation of resources, ensuring that network coverage and capacity can dynamically meet the fluctuating demands of users in crowded or rapidly changing scenarios. Moreover, where mobility patterns may vary unpredictably, such as in NNs, the ability to dynamically adjust network components enables seamless connectivity and optimal performance for users on the move. The various implementation forms within NNs place distinct demands on the new 6G network architecture, which must be addressed to fully realize the potential. Recommendation ITU-R M. 2160 [1] already covers scenarios for ubiquitous connectivity and introduces new metrics for coverage and sustainability. While NNs enable new applications and greater flexibility, integrating mobile components into the network introduces several challenges that require further research.

This paper examines the necessary architectural modifications and challenges in the scope of NNs and their importance in the context of the emerging 6G era. Section 2 reviews current literature and technological advancements in the scope of NNs, before the essential architectural requirements for integrating NNs into the 6G framework are outlined in Section 3. After introducing key demands, Section 4 proposes various architectural approaches to realize NNs in 6G. It discusses network slicing, deployment scenarios, orchestration frameworks, and furthermore identifies architectural challenges that arise from integrating NNs into existing frameworks. The paper concludes in Section 5 with a summary and an outline for future work.

## 2    Related Work

Numerous novel use cases, e.g. in manufacturing, agriculture, transportation, Programme Making and Special Events (PMSE) agriculture and Public Protection and Disaster Relief (PPDR) entail the need for spatial mobility. In addition, their requirements, such as data rates and latency, can often only be met by cellular systems. Key enabler for these use cases can be network slicing, which was introduced as part of 5G [2] and is also expected to be a cornerstone of 6G [3]. It is an architectural approach that enables the use of several independent virtual networks (slices) on a shared physical infrastructure. However, this requires an existing public network. As public networks are not always present, a reliable QoS cannot be guaranteed in mobile scenarios. Here, Non-Terrestrial Networks (NTNs) can improve connectivity of communication services. They contribute to extending coverage, enhancing capacity and ensuring connectivity in areas where terrestrial links may be temporarily limited or unavailable. Here, either direct or indirect access, via a backhauling link, can be granted [4]. Using direct access, the fulfillment of multiple Key Performance Indicators (KPIs) and efficiency can be an issue. This applies in particular to devices that are located spatially close to each other.

In addition, the possibility of private and self-operated cellular networks was introduced [5]. This allows a guaranteed coverage and less demands on the NTN link, as only the data network is connected via this link. In this context, the concepts of nomadic nodes [6] and temporary networks emerged [7]. For scenarios, where multiple base stations and a complete cellular network is mobile, the term NNs was introduced [8], whereas the combination with NPNs is promising.

Moreover, 6G systems are expected to be highly flexible or even show organic behavior [9]. In order to realize organic networks, adaptions on both Radio Access Network (RAN) and Core Network (CN) have to be made as demonstrated in [10]. In order to achieve organic CNs, key technologies, such as serverless [11] and live migration [12], are recently developed. On RAN side, cascaded networks can help to increase flexibility, as authors evaluated in [13]. In [14], the authors proposed so-called Dynamic Small Cells with Integrated Access Backhaul (IAB) using Centralized Unit (CU) - Distributed Unit (DU) split architecture. In IAB setups, which have been introduced in 3GPP Release 16, IAB nodes serve as relays linked to an IAB donor through a wireless backhaul connection [15]. This wireless link can utilize spectrum bands below and above 6 GHz, facilitating multi-hop operations. In this context, an IAB node may integrate User Equipment (UE) and DU functionalities, while the IAB donor is responsible for implementing CU and DU functionalities.

The introduction of dynamic network topologies increases concerns on trustworthiness of interactions. The exploration of trust relationships in the context of 6G extends beyond traditional security concerns. It encompasses the secure orchestration of resources across edge and cloud environments, and the assurance of privacy and integrity in the face of unprecedented data volumes [16]. As proposed in [17], communication based solutions to trust issues between peer networks can be introduced to provide a traceable ledger of past interactions without compromising the privacy of the actors. Mutual consensus on records can be used afterward for accountability of participants. In the case of NNs, this can be used for example as a channel for negotiations on spectrum usage.

## 3    Architectural Requirements for Nomadic Networks

To facilitate the integration of NNs, it is crucial to establish a holistic understanding of the requirements involved. By identifying and categorizing these requirements, we can streamline the design process and mitigate its inherent complexity. Our approach involved gathering a diverse set of requirements and identifying the dependencies between them. This analysis revealed a layered structure, wherein each layer of requirements stems from the implications of the preceding layer.

### 3.1    Key enablers for nomadic networks

In the first layer of requirements, the focus is on enabling the dynamic relocation of infrastructure components during runtime and what fundamental technical capabilities are needed to achieve this dynamicity.

#### 3.1.1    Functional split

The functional split between CU, DU, and Radio Unit (RU) can play a role in the architecture of NNs, in scenarios where not a complete network infrastructure is moving, but only parts of the RAN. The split could divide a base station into a fixed CU part and a moving DU part. A lower layer split between those would decrease costs, weight and resource consumption of the moving units, but in return would place increasing demands on the wireless and time variant midhaul link. A wireless backhaul link on the other hand would be required for full stack nomadic RAN nodes to be connected to a fixed CN.

### 3.1.2 Backhauling and resource allocation

In scenarios, where both RAN and CN are deployed on a moving infrastructure as a nomadic standalone system, the provisioning of an internet connection would require the N6 interface of the CN to be established via a wireless access point. The set of candidates for wireless access points is expected to get more diverse in the future, reaching from terrestrial links to drones, High Altitude Platform Stations (HAPSs), and satellite constellations [18]. This diversity of links is envisaged to improve network availability even in remote and hard-to-reach areas. Poor coverage zones and disaster areas could benefit from combining local nomadic nodes and high altitude access points, which could provide suitable network services without relying on local infrastructure.

Another kind of resources to be used by NNs are compute resources for CN services. In the event that the NN uses the existing on-premises infrastructure to deploy its own CN and edge computing applications, the resources (compute and power) must be purchased from infrastructure providers. In the case of a moving network with hard requirements regarding the latency between CN and RAN but no possibility of using an integrated CN, the software deployment needs to be dynamically migrating to always reside in proximity to the RAN. This requires protocols and interfaces for information exchange between infrastructure providers and operators for dynamic allocation of compute resources at runtime.

## 3.2 Management and control of nomadic 6G components

Considering the mobility of networks, the implications for management and control of network components build the second layer of requirements.

### 3.2.1 Interoperability

Looking at the continuous roll-out of new generations, the dynamic placement of RAN nodes in most nomadic use cases, spectrum limitations and the dynamic network topology envisioned for 6G systems, interoperability with network parts of previous generations becomes important for comprehensive use of nomadic systems. In consequence, NNs can be expected to regularly come in contact with 5G systems and therefore are required to stay backwards compatible, with legacy systems either being aware of the nomadic nature of some new network parts or not. Therefore, compatibility to existing management and virtualization frameworks is of high interest when identifying the architectural demands of NNs. By describing NomadicNetwork-as-a-Service (NNaaS), which can be provided by different kinds of Mobile Network Operators (MNOs), identified technical gaps can be mapped to management procedures and Network Functions (NFs). The increasing flexibility on the infrastructure layer could enhance the scope of Network Slicing as a Service (NSaaS) by providing the foundation for virtual networks over both fixed and moving infrastructures.

The ability of the network to adapt even physical deployment of nodes to external conditions enhances the overall flexibility and reduces resource consumption. The possibility of dynamic fallback from a fully connected network to local reduced capability networks increases resilience, at least for local connectivity and service deployments. This comes at the cost of higher complexity regarding network management, as dynamic allocation and deployment of NFs requires formalization of possible combinations and network capabilities, as well as implementation of redeployment procedures at runtime.

### 3.2.2 Exchange and collaboration

Derived use-cases from Lindenschmitt et al. require certain nomadic features as well as technical capabilities [8]. Likewise the nomadic features require technical tools for the involved processes across CN, RAN and stakeholders. Multiple stakeholders are potentially involved in a nomadic scenario due to the variance in which they can be set up. As multiple distributed components and distinct stakeholders can be collaborating, additional logical challenges arise regarding permissions, identity, confidentiality and trust. The segmentation of the network components involved means that both the behavior-determining software and the executing hardware can each be contributed by a different party. This results in the need for explicit coordination between the stakeholders involved:

*Platform Provider*: The party that provides an execution environment for software services. This includes the executing hardware as well as potentially a (software-defined) management platform for orchestrating the execution. This can be for example a cloud infrastructure provider or operator of any (server) infrastructure.

*Infrastructure Service Operator*: The party to run the services defining the CN and parts of the RAN. *Nomadic Network Provider*: The party enabling connectivity as NNaaS introduced in Section 3.2.1, i.e. operating one of the nomadic RANs scenarios described in Section 4.2. *End user*: Everyone not contributing to the infrastructure but utilizing it. Essentially any application using connectivity.

Scenarios involving the possibility of radio interference with adversary RAN nodes, as presented for example in [19], will require overhead communication between operators for radio resource allocation. The highest extent of overhead is required, if two interfering networks are operated by stakeholders which are completely independent of each other and if the route of the moving network parts is not known beforehand. If interference situations can be forecasted, e.g., when a new RAN node spawns at a fixed location or moves on a predefined path, radio resource allocation can be negotiated beforehand, but there still need to exist procedures for setting up the modified resource allocation when the operation of the new node starts. Only in scenarios where all interferences with adversary nodes can be neglected, e.g., disaster areas where no network is available, no overhead for radio resource negotiation is necessary.

By attaching to an already provisioned CN, nomadic RAN nodes can be used to enhance the network's coverage in a specific area. In this scenario, subscriber management is provided by the MNO in charge of the CN, while the nomadic RAN not necessarily needs to be managed by the same operator. So, instead of communicating with infrastructure providers, the NN needs to interact with MNO functions for dynamic topology reconfiguration. The mobile nature of some nomadic nodes requires routing mechanisms for the setup procedure to be independent of the location.

## 3.3 Ensuring trustworthiness of stakeholders

With networks in motion and active interactions among all stakeholders, the implications arising for the relationships between these interacting entities are the starting point for the third layer of requirements.

Section 3.2 points out, that realistic scenarios for NNs rely on information exchange for session setup and management between many different stakeholders involving MNOs, infrastructure providers, service providers, or end users. Dynamic business relations and granular billing concepts on a service level can be expected to require a high degree of automation on the management layer. Object to such business relations can be any physical or virtual resource necessary to compose end-to-end connectivity of the required quality [20].

The seamless interaction of network components requires that the systems not only provide each other with encapsulated services, but also allows access to open interfaces. From the perspective of end users in public networks, composite connectivity solutions result in the data plane being handled by multiple stakeholders, which might change over time due to the mobility of nomadic nodes.

On the other hand, by enabling the option of deploying a fully integrated CN and RAN on privately owned moving infrastructure, control over all NFs and resources is handed over to the owner/operator of the NN. In this scenario, the trustworthiness of all internal services is increased, as the user is also the operator of the network. Critical trust relations in this setup are considered when theNN needs to interact with adversary networks, for example in co-existence scenarios, or when setting up a wireless N6 link. In nomadic scenarios, such situations can occur not only at setup time, but also during operation. An automatable framework for identity management and authentication is required, not only for end devices, but for all network components involved.

# 4 Approaches on a fundamental architecture for 6G Nomadic networks

After defining the architectural requirements for NNs, the following section provides an implementation of these in the 6G framework. Various strategies for the integration of NNs in network slicing are considered, different deployment scenarios are discussed and orchestration frameworks are presented.

## 4.1 Nomadic Networks in Sliced 6G Network

Concerning the essential features of NN such like constrained access, limited life-cycle, and resource isolation, it is hard to disregard its correlation with the concept of network slicing. Indeed, as we have summarized in Section 2, network slicing provides a convenient approach to realize spawned RAN and migrated CN, without excluding other scenarios.

While implementing and managing NNs on the network slice instance layer seems a straightforward solution, it lacks of flexibility and efficiency. First, the users in a NN may be polymorphic and involved in various use cases (e.g., enhanced Mobile Broad-band (eMBB) and Ultra Reliable Low Latency Communication (uRLLC)), which cannot be implemented within the same network slice. Second, in the case of time-constrained NNs, e.g. spawned RAN, and virtually implemented steered RAN, repeatedly turning a slice on-and-off can be complex. The necessary introduction of an "idle" status of a NN slice is bringing extra complexity to not only the real-time inter-slice resource allocation, but also the Service Level Agreement (SLA) design and slice life-cycle management.

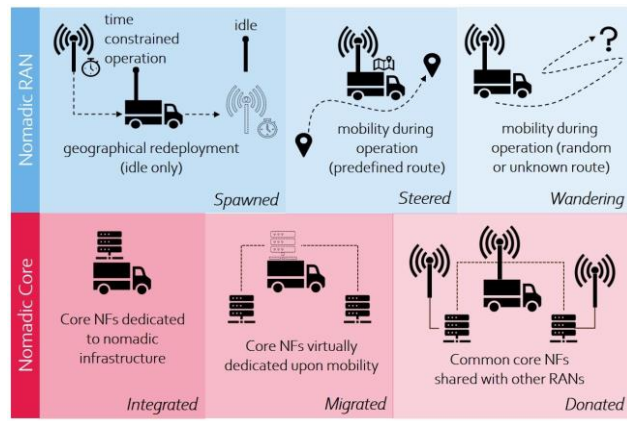

Figure 1: Deployment options of nomadic RAN and CN

Therefore, we propose that NNs shall be implemented and managed on the service instance layer. This first enables the support heterogeneous use cases within a single NN, by assigning multiple network slices of various types to it. It also simplifies the implementation of time-constrained NNs, where the turn-on and turn-off of a NN can be realized by adding network slices to it, and terminating its dedicated network slices, respectively. Furthermore, understanding every NN as a service instance is leading us towards a potential novel business paradigm of NNaaS.

## 4.2 Multiple Deployment Scenarios

Primarily focusing on the issue of dynamic radio access, a NN is essentially implementing its dedicated RAN. Generally, NNs are dynamic over time, i.e., they can be flexibly turned on-and-off. Depending on the dynamics over space, there are generally three deployment scenarios for a nomadic RAN:

*Steered*: The nomadic RAN moves along a predefined route. In this scenario, the MNO can anticipate the potential overlap of the nomadic RAN with other RANs that are either static, spawned, or steered. As a straightforward solution, a steered RAN can be implemented with its dedicated moving network infrastructure, e.g. Automated Guided Vehicles (AGVs), Unmanned Aerial Vehicles (UAVs) or High Altitude Platforms (HAPs). Alternatively, a steered RAN can also be virtually implemented, by sequentially turning on/off a set of spawned RANs that are identically or similarly specified and located along the route.

*Wandering*: The nomadic RAN moves without a predefined route, or the predefined route is unknown by the network. In this case, a precise anticipation of potential overlap with other RANs will be impossible. This scenario is mostly concerned with dynamic movement of network infrastructure, where the wandering is spontaneously realized. A virtual implementation, by means of invoking spawned RAN along the dynamic wandering path, is theoretically possible. However, it is hardly feasible in practice, due to the lack of time for activating the next spawned RAN on the wandering path.

Upon the logical dedication and physical deployment of the NFs, there are three deployment options for a nomadic CN: *Integrated*: A complete set of CN NFs are deployed within the nomadic infrastructure as a dedicated part, and with dedicated resources. This option grants not only the maximal independence and flexibility, but also advantages in security and reliability. However, its high Capital Expenditures (CAPEX) and Operating Expense (OPEX) becomes an inevitable drawback.

*Migrated*: The NN has its own logically independent core NFs, which are physically deployed on the static infrastructure. When the nomadic RAN nodes move, the nomadic CN NFs can be correspondingly redeployed. This option can be implemented through CN slicing, where infrastructural resources are dynamically allocated upon the mobility of the NN.

*Donated*: The NN does not contain any core NF, but only RAN functions. The nomadic RAN is always attached to an existing CN, sharing it with other RANs. From the perspective of operation, the options of donated and migrated nomadic CN distinguish from each other primarily by the ownership and administrative permissions of the network. Since the NN concept is primarily about the mobility of infrastructure, we can refer to the options of steered RAN and wandering RAN cumulatively as *moving RAN*, which is opposite to the geographically fixed spawned RAN. Similarly, the options migrated and donated CN can be concluded as *ex-sourced CN*, which is externally sourced from the static infrastructure, in opposition to the integrated CN with the capability of mobilizing its physical infrastructure.

The deployment options of CN and RAN, as illustrated in Figure 1 and detailed in Figure 2, can be flexibly combined, in order to fulfill different requirements of various use scenarios:

- The combination of an integrated CN and a moving RAN aims at deploying the entire network on a mobilized infrastructure, so that both the functionality and the isolation from external networks can remain complete throughout the operation of a moving NN. Typical scenarios include container ships, police and military networks, etc.
- The combination of an integrated CN and a spawned RAN enables agile deployment of a complete network at a certain location, which is often helpful in areas where no reliable backhaul of the Public Local Mobile Network (PLMN) is possible or desired. Typical scenarios of this combination are PPDR or agriculture.
- The combination of an ex-sourced CN and a moving RAN is capable of providing a mobilized coverage of radio access at low implementation cost, when the independence and isolation of the core NFs is not crucial. For example, IoT applications within a truck platoon or an UAV swarm can benefit from this combination.
- The combination of an ex-sourced CN and a spawned RAN is basically a temporary network slice to deliver mobile service within limited area and time period.

More specifically, when using a moving RAN, the choice between the steered and migrated options is upon the mobility model of the RAN. On the other hand, when an exsourced CN shall be deployed, the subtle choice between migrated and donated CNs depends on the interests of different stakeholders in the data and the NF.

## 4.3 Nomadic Network Orchestration Framework

The efficient deployment of NNs depends on their ability to integrate with existing frameworks (both at network and management levels) established by leading SDOs like 3GPP, ETSI, and O-RAN. In the following, we discuss this integration for key aspects of NN operation: mobility control, resource scheduling, and life-cycle management.

interfaces provide a standardized framework for life-cycle management. NNs can benefit from established procedures for deployment, configuration, monitoring,

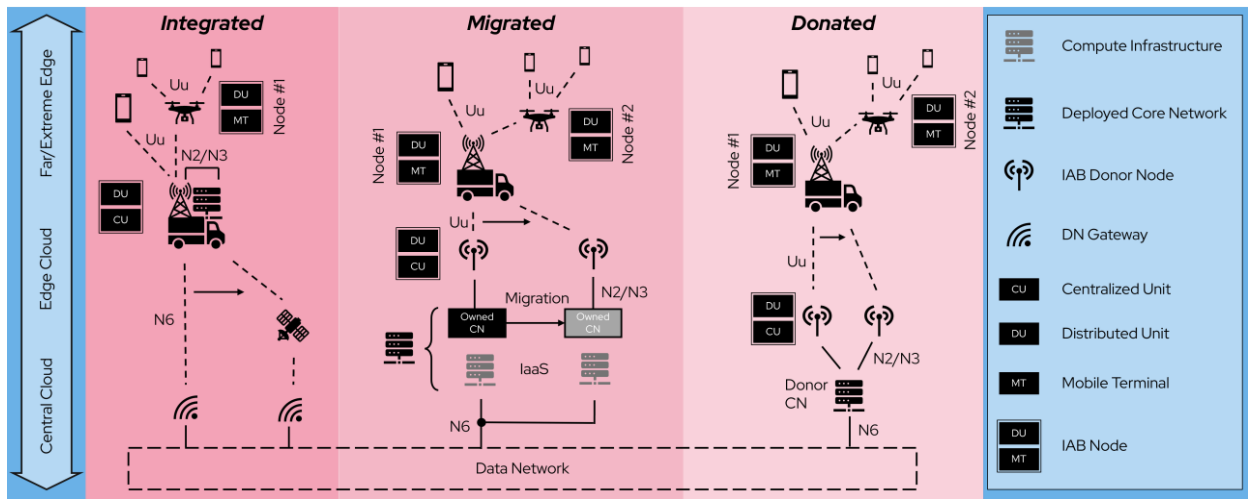

Figure 2: Interfaces and network components for different deployment options of the CN

For mobility control, NNs can leverage the established infrastructure within these frameworks. In 3GPP architecture, NN controllers can register with existing Access and Mobility Management Function (AMF) similar to UE registration. This allows the AMF to track the location and connectivity status of NN elements, ensuring smooth handovers and maintaining network continuity. Similarly, in the context of ETSI's Multi-Access Edge Computing (MEC) standards, NN controllers can act as MEC applications. They can leverage existing mechanisms for registration, discovery, and resource management, facilitating their integration into existing MEC deployments.

Moving on to resource scheduling, NN can benefit from established interfaces for resource allocation. The 3GPP Next Generation Radio Access Network interface, especially the N2 or N3 interface as illustrated in the different deployment options for CNs in Figure 2, can provide a communication channel for NN controllers to negotiate resource allocation with the network scheduler. This allows them to request specific bandwidth or radio resources for their operation. Additionally, O-RAN's Open Fronthaul Interface (O-FH) and Open Fronthaul Interfaces Alliance (OFIA) specifications offer vendor-neutral control of resource allocation. NN elements can leverage these specifications to integrate with different RAN vendors' equipment, promoting interoperability within the network.

Life-cycle management of NNs can also be addressed by utilizing existing mechanisms. Both 3GPP and ETSI define protocols for network entity registration and discovery. NN elements can register with the network upon deployment and advertise their capabilities. This allows the CN to track their presence, manage their resources, and ensure proper operation. Furthermore, O-RAN's Open Operations, Administration, and Maintenance (OAM) and decommissioning from OAM. This ensures smooth integration into existing network management practices.

## 4.4 Architectural Challenges

While the potential benefits of NNs are substantial, integrating them into existing frameworks presents several architectural challenges. Firstly, the dynamic nature of NNs clashes with the relatively static configurations of traditional mobile networks. Existing frameworks are optimized for managing a fixed set of network elements, with established handover procedures and resource allocation strategies. NNs, with their constant movement and changing capabilities, require novel approaches to ensure seamless handovers, efficient resource allocation, and consistent network performance. This necessitates the development of flexible and adaptive mechanisms within the existing frameworks to accommodate the dynamic nature of nomadic deployments. Secondly, security considerations pose a significant challenge. NNs introduce new attack vectors into the network due to their mobility and potential for unauthorized access. Existing security mechanisms may not be sufficient to efficiently address these challenges. Integrating nomadic elements necessitates extending authentication and authorization protocols to ensure secure communication and prevent unauthorized access to the underlying network resources. Additionally, robust mechanisms for dynamic trust establishment and secure handover procedures are crucial to safeguard network integrity and user privacy. Addressing these security challenges will be paramount for ensuring the secure and reliable operation of NNs within existing frameworks.

## 5  Conclusion

This paper has highlighted the potential of NNs for a future 6G standard. NNs can overcome the limitations of stationary networks by enabling flexible and comprehensive connectivity solutions in various scenarios such as PMSE, PPDR and mobile industrial applications. The integration of NNs into a 6G infrastructure requires significant architectural changes, including dynamic resource allocation, seamless handover mechanisms and enhanced security protocols. The importance of interoperability with existing networks and the use of NTNs to extend coverage has also been discussed. Multiple deployment scenarios of NN for CN and RAN were introduced and subsequently put into context with applications and stakeholders. An orchestration framework for NN based on the N2 interface was discussed and challenges for the architecture were identified.

Future work should focus on developing flexible mechanisms for resource allocation, ensuring secure interactions between stakeholders, and refining the orchestration of NFs to support the dynamic nature of NNs. By addressing these challenges, the capabilities of NNs can be fully utilized so that future 6G networks can meet the evolving needs of different sectors, including industry, manufacturing, agriculture and public services.

## Acknowledgment

The authors acknowledge the financial support by the German *Federal Ministry for Education and Research (BMBF)* within the projects »Open6GHub« {16KISK003K & 16KISK004} and »6GTakeOff« {16KISK067}.